# Beyond Information Retrieval: Generative AI as an Epistemic Arbiter to Enhance Collaborative Problem-Solving

Jiaxin Zou[1][0009-0009-2327-3208], Xiaoming Zhai[2][0000-0003-4519-1931], Chunlei Gao[1][0009-0009-7006-293X]

[1] School of Education, Jiangxi Normal University, 99 Ziyang Ave., Nanchang 330022, Jiangxi, China
[2] University of Georgia, Athens, GA 30666, USA
zjxjannet@163.com
Xiaoming.Zhai@uga.edu
gaochunlei@jxnu.edu.cn

**Abstract.** Generative AI (GAI) creates new opportunities for collaborative problem-solving (CPS), yet its role in shaping student interaction remains unclear. To address this gap, we conducted a six-week quasi-experimental study with 201 fifth-grade students in two conditions: with and without GAI. Chi-square analysis showed significant differences in CPS behavior distributions between groups. Compared with the control group, the GAI-supported group demonstrated more social behaviors, particularly engagement and conflict management, but less frequent cognitive behaviors such as task planning and solution reasoning. Lag sequential analysis further revealed distinct interaction patterns: while the control group followed a more conventional transition from listening to planning, the GAI group showed a robust pathway from task planning to conflict management to solution reasoning. Thematic analysis of AI interaction logs suggested that students used GAI as an epistemic arbiter, drawing on AI-generated facts and visualizations to resolve disagreements constructively. These findings suggest that GAI reshapes CPS by mediating the transition from social conflict to collaborative reasoning.

## 1 Introduction

Collaborative problem-solving (CPS) is recognized as a pivotal 21st-century competency, necessitating the ability to share knowledge, negotiate perspectives, and co-construct solutions [11, 17]. While traditional instruction often struggles to foster these complex skills due to passive pedagogical structures [19], the advent of generative AI (GAI) offers transformative potential [21]. Despite the research on GAI in education, there are dimensions that warrant further investigation. First, research has predominantly focused on individual learning scenarios, such as knowledge transmission and personalized tutoring [10], leaving the dynamics of AI-supported group collaboration insufficiently explored. Second, current research on student-GAI interaction predominantly concentrates on higher education [9], with limited attention paid to elementary students. Third, while existing research enriches our understanding of CPS [13], many studies rely on static analyses that overlook the temporal flow of interaction. To address these gaps, this study investigates how GAI influences elementary students' CPS using a mixed-methods design. Drawing on classroom video and GAI interaction logs, we apply lag sequential analysis (LSA) and thematic coding to examine (1) differences in CPS behavior distributions between groups with and without GAI support, (2) their respective sequential patterns of CPS, and (3) the functional roles of GAI in mediating and shaping CPS micro-dynamics.

## 2 Related Work

CPS is an important educational goal that requires teamwork, negotiation, and shared knowledge construction [17]. In ATC21S, CPS refers to approaching a problem responsively through collaboration and idea exchange [5]. In recent years, researchers have concentrated on the micro-interaction processes of CPS. Prior studies have shown that learning in CPS is embedded not only in outcomes but also in real-time interaction processes, especially in how groups manage coordination and conflict. However, effective CPS implementation in elementary education is in challenge. Young learners' social-emotional skills are in the process of development and thus limited, often encounter difficulties with "free-riding," lack of coordination, or engaging in conflicts [19]. Although traditional scaffolds can provide structural support, they often remain insufficiently responsive to moment-to-moment group dynamics [4]. With the introduction of GAI, CPS increasingly takes place in human-AI collaborative systems rather than in peer-only interactions. Existing studies suggest that GAI may influence communication, division of labor, and knowledge work within collaborative settings [16]. Most prior research has focused on higher education or broad interaction categories, providing limited insight into how GAI shapes the temporal micro-dynamics of collaboration [3]. Notably, there is a significant lack of fine-grained empirical evidence regarding how students in the period of cognitive development in elementary education organically incorporate the outputs of GAI into their social conflict management and collaborative reasoning cycles. This study employs LSA to decode the behavioral sequence patterns in classroom videos.

It utilizes thematic analysis to conduct an in-depth interpretation of the discourse logic within interaction logs. This study draws on Distributed Cognition Theory, which conceptualizes cognition as distributed across people, tools, and social interaction [8], together with the ATC21S framework for distinguishing social and cognitive dimensions of CPS [7]. This combined perspective supports an examination of how GAI mediates elementary students' CPS processes and reshapes the relationship between social interaction and collaborative reasoning.

# 3 Methods

## 3.1 Participants

Participants were 201 fifth-grade students (101 boys, 100 girls; mean age = 11) from four classes in a primary school. The classes were assigned to either an experimental group (EG) with GAI or a control group (CG) without GAI. Students worked in 40 small groups of five to six members. The study followed institutional ethical guidelines, and all data were anonymized.

## 3.2 Experimental Procedure and Data Analysis

A six-week quasi-experiment was conducted, with one 80-minute session per week. Before the intervention, students completed a CPS skills test adapted from Von Davier et al. [15], which confirmed no significant difference in prior CPS levels between groups ($F = .001$, $p = .828$). During the six-week experiment, students collaborated to complete six tasks. The elementary science curriculum, named "Interstellar Navigation: Reaching for the Sky," was developed based on Project Based Learning principles and comprised six sequential modules. The only difference was that the EG used Nano AI as the GAI tool during collaboration, whereas the CG relied on printed materials. Both groups were taught by the same instructor.

To analyze CPS processes, classroom videos were segmented into 15–20 second units, yielding 1,154 behavioral entries across 300 minutes of video. Based on a CPS coding framework adapted from Sun et al. [7], two independent researchers coded each unit. CPS behaviors were coded into four social categories (participation, listening and responding, conflict management, responsibility sharing) and four cognitive categories (task planning, resource management, solution reasoning, monitoring and correction) [7]. Inter-rater reliability was established on 20% of the data, with a Cohen's kappa of .717. To address the first research question, chi-square tests were used to compare the distribution of CPS behaviors between the two conditions, with phi ($\varphi$) reported as the effect size. To examine sequential patterns, LSA was conducted using GSEQ. Adjusted residuals greater than 1.96 were treated as statistically significant. To address the third research question, thematic analysis was applied to the 364 GAI interaction logs.

# 4 Results

## 4.1 Different Characteristics of CPS Behavior Patterns Between the Two Groups

To examine the impact of GAI on students' CPS processes, we compared the distribution of CPS behaviors between the EG and the CG. The frequencies and percentages are presented in Table 1. The EG showed a significantly higher proportion of social behaviors than the CG (73.137% vs. 63.605%; $\chi^2 = 12.122$, $p < .001$, $\varphi = .103$), particularly in participation (45.581% vs. 39.341%; $\chi^2 = 4.597$, $p = .032$, $\varphi = .063$) and conflict management (4.506% vs. 1.733%; $\chi^2 = 7.340$, $p = .007$, $\varphi = .080$). In contrast, the CG showed more cognitive behaviors, especially task planning (13.692% vs. 9.879%; $\chi^2 = 4.034$, $p = .045$, $\varphi = .059$), solution reasoning (6.066% vs. 1.560%; $\chi^2 = 15.973$, $p < .001$, $\varphi = .118$), and monitoring and correction (8.146% vs. 5.026%; $\chi^2 = 4.564$, $p = .033$, $\varphi = .063$). These results suggest that GAI shifted CPS behavior toward social engagement, while the control condition showed more frequent overt cognitive behaviors. Effect sizes were small overall.

**Table 1.** Frequency and percentage of CPS behavior in the two conditions.

| Behavior | EG | CG | $\chi^2$ | $p$ | φ |
|---|---|---|---|---|---|
| | Frequency (%) | Frequency (%) | | | |
| Social skills | 422 (73.137%) | 367 (63.605%) | 12.122 | .000 | .103 |
| Participation | 263 (45.581%) | 227 (39.341%) | 4.597 | .032 | .063 |
| Listening and responding | 53 (9.185%) | 62 (10.745%) | .782 | .376 | |
| Conflict management | 26 (4.506%) | 10 (1.733%) | 7.340 | .007 | .080 |
| Responsibility sharing | 80 (13.865%) | 67 (11.612%) | 1.317 | .251 | |
| Cognitive skills | 155 (26.863%) | 210 (36.395%) | 12.122 | .000 | .103 |
| Task planning | 57 (9.879%) | 79 (13.692%) | 4.034 | .045 | .059 |
| Resource management | 60 (10.399%) | 50 (8.666%) | 1.005 | .316 | |
| Solution reasoning | 9 (1.560%) | 35 (6.066%) | 15.973 | .000 | .118 |
| Monitoring and correction | 29 (5.026%) | 47 (8.146%) | 4.564 | .033 | .063 |

### 4.2 Differences in the CPS Patterns Between the Two Groups

LSA was conducted to examine differences in CPS behavioral patterns between the two groups, with adjusted residuals above 1.96 treated as statistically significant. The EG showed a coherent cross-dimensional pattern characterized by a cognitive-social-cognitive sequence. Specifically, there was a significant transition from “task planning” (C1) to “conflict management” (S3) ($Z = 2.29$, $p < .05$), followed by a stronger transition from “conflict management” (S3) to “solution reasoning” (C3) ($Z = 3.90$, $p < .01$). This continuous pathway (C1→S3→C3) suggests that GAI-supported collaboration helped transform disagreement into solution-oriented reasoning. The CG showed a more fragmented pattern without a continuous behavioral chain. Its most prominent transition was from “listening and responding” (S2) to “task planning” (C1) ($Z = 3.02$), while “responsibility sharing” (S4) led separately to “resource management” (C2, $Z = 1.99$) and “solution reasoning” (C3, $Z = 2.66$). Overall, the EG demonstrated a more integrated CPS sequence, whereas the CG showed relatively isolated transitions from social to cognitive behaviors.

### 4.3 Functional Roles of GAI in Supporting the CPS Process

To further examine the micro-mechanisms underlying the observed behavioral transitions, a thematic analysis was conducted on 364 GAI dialogue logs. The analysis identified main themes of the use of GAI, relevant descriptions are presented in Table 2.

Students most frequently used GAI for knowledge acquisition (n = 220, 60.440%), particularly to obtain factual information (n = 85) and basic conceptual explanations (n = 75). They also used GAI for visualization requests (n = 90, 24.725%) and analytical support (n = 30, 8.242%). These findings suggest that students used GAI not merely as an information provider, but as an external source of evidence and representational support that helped them resolve disagreement and advance collaborative reasoning.

**Table 2.** Coding scheme and samples of student-GAI interaction logs in the CPS process.

| Theme | Subtheme | Description |
|---|---|---|
| Knowledge acquisition | Foundational concepts | Seeking basic definitions or introductory explanations. |
| | Factual data | Seeking specific facts or numerical information. |
| | Comparative analysis | Comparing characteristics of different objects. |
| | Mechanisms & principles | Exploring the underlying causes or mechanisms. |
| Visualization requests | Thematic images | Requesting images based on a given theme. |
| | Drawing guidance | Seeking instructions or techniques for drawing. |
| | Stylized generation | Specifying artistic styles for generated content. |
| | Iterative refinement | Refining outputs or requesting improvements. |
| Analytical support | Data organization | Requesting structured data tables. |
| | Thought organization | Requesting diagrams or structured frameworks. |
| | Report generation | Requesting analytical reports. |
| Solution-oriented guidance | Procedural steps | Asking for step-by-step procedures. |
| | Design solutions | Requesting design drawings or proposals. |
| | Materials & methods | Inquiring about the materials and methods to use. |
| Irrelevant queries | Off-topic | Deviating from the course learning objectives. |
| | Test queries | Testing the AI's capabilities or boundaries. |

# 5 Discussion

This study investigated the impact of GAI on the CPS processes of students by analyzing behavioral interaction patterns under two distinct learning conditions. The findings reveal differences in CPS processes between the two groups. Compared with the CG, the EG showed higher levels of participation and conflict management, whereas the CG demonstrated more frequent task planning and solution reasoning. This pattern suggests that the integration of GAI shifted CPS behavior toward the social dimension. Rather than functioning merely as a tool, GAI appeared to reshape the socio-cognitive structure of collaboration by lowering the threshold for participation and introducing new occasions for negotiation [14]. At the same time, the lower frequency of overt cognitive behaviors in the EG suggests that some planning and reasoning tasks may have been offloaded to the AI, raising concerns about reduced opportunities for critical and reflective thinking without appropriate pedagogical support [20].

The LSA results further highlight a key difference in CPS dynamics between the two groups. The CG showed a fragmented and linear path, with transitions such as listening and responding to task planning, indicating a more conventional collaboration pattern [1]. While the EG developed a robust sequence from task planning to conflict management to solution reasoning. This pattern suggests that GAI did not simply reduce conflict, but helped transform disagreement into collaborative reasoning. In this sense, GAI functioned as an “epistemic arbiter,” providing evidence that supported students in resolving disputes constructively and moving toward higher-order reasoning [6, 18]. The thematic analysis provides further support for this interpretation. Students primarily used GAI for knowledge acquisition, visualization, and analytical support, suggesting that GAI functioned as an epistemic scaffold rather than a simple answer generator. By offering factual information and representational support, GAI helped students externalize ideas and stabilize shared understanding [2, 12]. However, the logs also suggest a potential risk of cognitive reliance, as students rarely questioned or critically evaluated AI-generated outputs. This finding indicates that GAI-supported CPS should be accompanied by explicit pedagogical scaffolds that prompt students to examine, compare, and justify AI-generated evidence rather than accept it uncritically.

## 6 Conclusion

This study examined how GAI mediated CPS among fifth-grade students. The results showed that, compared with the control group, GAI-supported groups engaged more in participation and conflict management but less in overt task planning and solution reasoning. Importantly, LSA revealed a robust transition from task planning to conflict management to solution reasoning in the GAI condition. These findings suggest that GAI did not simply provide information, but functioned as an epistemic arbiter that helped students use AI-generated facts and visualizations to resolve disagreement and advance collaborative reasoning.

**Acknowledgments.** This work was supported by the National Education Science Planning Project [BSA250242].

**Disclosure of Interests.** The authors have no competing interests to declare.

## References

1. Alexander, R.J.: Developing dialogic teaching: Genesis, process, trial. Research Papers in Education **33**(5), 561–598 (2018)
2. Arcolin, C., Schaefer, W., Matos, F., et al.: Use of Generative Artificial Intelligence in an Introductory Science Course. Technology, Knowledge and Learning (2025)
3. Feng, S.: Group interaction patterns in generative AI-supported collaborative problem solving: Network analysis of the interactions among students and a GAI chatbot. British Journal of Educational Technology **56**(5), 2125–2145 (2025)
4. Fraser, B.J.: The Evolution of the Field of Learning Environments Research. Education Sciences **13**(3), 257 (2023)

5. Griffin, P., McGaw, B., Care, E.: Assessment and Teaching of 21st Century Skills. Springer, Netherlands (2018)
6. Helal, M.Y.I., Elgendy, I.A., Albashrawi, M.A., et al.: The impact of generative AI on critical thinking skills: a systematic review, conceptual framework and future research directions. Information Discovery and Delivery (2025)
7. Hesse, F., Carew, E., Buder, J., Sassenberg, K., Griffin, P.: A framework for teachable collaborative problem solving skills. In: Griffin, P., Carew, E. (eds.) Assessment and teaching of 21st century skills, pp. 37–56. Springer, Dordrecht (2015)
8. Hutchins, E.: Cognition in the wild. The MIT Press, Cambridge (1995)
9. Kim, J., Lee, S., Detrick, R., Wang, J., Li, N.: Students-Generative AI interaction patterns and its impact on academic writing. Journal of Computing in Higher Education (2025)
10. Lee, Y.: Developing a computer-based tutor utilizing Generative Artificial Intelligence (GAI) and Retrieval-Augmented Generation (RAG). Education and Information Technologies **30**(6), 7841–7862 (2025)
11. OECD: PISA 2015 Assessment and Analytical Framework: Science, Reading, Mathematic, Financial Literacy and Collaborative Problem Solving, revised edn. OECD Publishing, Paris (2017)
12. Owen, A.E., Roberts, J.C.: Visualisation Design Ideation with AI: A New Framework, Vocabulary, and Tool. Future Internet **16**(11), 406 (2024)
13. Shin, Y., Jung, J., Choi, S., Jung, B.: The influence of scaffolding for computational thinking on cognitive load and problem-solving skills in collaborative programming. Education and Information Technologies **30**(1), 583–606 (2024)
14. Shloul, T.A., Mazhar, T., Abbas, Q., et al.: Role of activity-based learning and ChatGPT on students' performance in education. Computers and Education: Artificial Intelligence **6**(3), 100219 (2024)
15. Von Davier, A.A., Hao, J., Liu, L., Kyllonen, P.: Interdisciplinary research agenda in support of assessment of collaborative problem solving: lessons learned from developing a Collaborative Science Assessment Prototype. Computers in Human Behavior **76**, 631–640 (2017)
16. Wang, T., Long, T., Mei, A., Li, J., Liu, Z.: Fusing Six-Hat thinking with AI: How the Green Hat and ChatGPT Co-Regulate pre-service teachers' instructional design—insights from epistemic network analysis. Thinking Skills and Creativity **59**, 102025 (2025)
17. Wei, X., Wang, L., Lee, L., Liu, R.: The effects of generative AI on collaborative problem-solving and team creativity performance in digital story creation: an experimental study. International Journal of Educational Technology in Higher Education **22**(1), 23 (2025)
18. Xie, K., Miller, N.C., Allison, J.R.: Toward a social conflict evolution model: Examining the adverse power of conflictual social interaction in online learning. Computers & Education **63**, 404–415 (2013)
19. Zahra, I., Neo, M., Hew, S.H.: Level-Up Learning with CIGLE Framework: Enhancing Learning Through Interactive, Game-Based Collaboration for Effective Problem Solving. International Journal of Technology **16**(1), 187–206 (2025)
20. Zhang, W., Liu, X.: Artificial Intelligence-Generated Content Empowers College Students' Critical Thinking Skills: What, How, and Why. Education Sciences **15**(8), 977 (2025)
21. Zheng, L., Shi, Z., Gao, L.: A generative artificial intelligence-enhanced multiagent approach to empowering collaborative problem solving across different learning domains. Computers & Education **241**, 105489 (2025)